\documentclass[preprint,3p,11pt,sort&compress]{elsarticle}

\usepackage{amsmath}
\usepackage{amssymb, bm}
\usepackage{hyperref}
\usepackage{lipsum}
\usepackage{xcolor}
\usepackage{lineno}
\usepackage{setspace}
\usepackage{upgreek}
\usepackage{booktabs}

\journal{journal}

\begin{document}

\begin{frontmatter}



\title{On decision-theoretic model assessment for structural deterioration monitoring}


\author[NTUA]{Nicholas E. Silionis}
\author[NTUA]{Konstantinos N. Anyfantis\texorpdfstring{\corref{cor}}{}}

\affiliation[NTUA]{organization={Ship-Hull Structural Health Monitoring (S-H SHM) Group, School of Naval Architecture and Marine Engineering, National Technical University of Athens},
            addressline={9 Heroon Polytechniou Av.}, 
            city={Athens},
            postcode={15780 Zografos}, 
            country={Greece}}

\ead{kanyf@naval.ntua.gr}
\cortext[cor]{Corresponding author. Tel.: +30 210 772 1325}

\begin{abstract}
As data from monitored structures become increasingly available, the demand grows for it to be used efficiently to add value to structural operation and management. One way in which this can be achieved is to use structural response measurements to assess the usefulness of models employed to describe deterioration processes acting on a structure, as well the mechanical behavior of the latter. This is what this work aims to achieve by first, framing Structural Health Monitoring as a Bayesian model updating problem, in which the quantities of inferential interest characterize the deterioration process and/or structural state. Then, using the posterior estimates of these quantities, a decision-theoretic definition is proposed to assess the structural and/or deterioration models based on (a) their ability to explain the data and (b) their performance on downstream decision support-based tasks. The proposed framework is demonstrated on strain response data obtained from a test specimen which was subjected to three-point bending while simultaneously exposed to accelerated corrosion leading to thickness loss. Results indicate that the level of \textit{a priori} domain knowledge on the deterioration form is critical.
\end{abstract}




\end{frontmatter}


\section{Introduction} \label{Seq1}
Recent advances in monitoring technology have increased the availability of data obtained from engineered systems \citep{Rogers2020, Bull2023, Mieloszyk2021}, which can be leveraged to create added value for their operation and management (O\&M). A typical example of this can be found in Structural Health Monitoring (SHM), where gathered structural response data (e.g., strain or acceleration) are transformed into decisions regarding the health state of the structure in question. These decisions are typically categorized hierarchically into different SHM-related tasks \citep{Rytter1993}; namely determining the existence of damage (diagnosis), its type, location and/or extent (identification) and predicting its remaining useful life (prognosis) \citep{Farrar2012}.

Whereas the first two tasks can be treated equally effectively using purely data-driven \citep{Bull2018, Svendsen2022, Jones2022}, as well as model-based SHM approaches \cite{Behmanesh2015, Behmanesh2017, Argyris2018}, a case can be made that for prognostic applications, which feature prominently in the predictive maintenance approach to O\&M, model-based methods are oftentimes better suited \citep{Kamariotis2024}. The reason lies in the inherent inability of data-driven models to extrapolate beyond their training horizon, which hampers their ability to provide predictions on the evolution of structural deterioration and thus make the transition from diagnosis to prognosis \citep{Farrar2012}. However, that is not to say that data-driven methods cannot be used effectively for prognostics; indeed they can, with the caveat that they require training sets that contain information corresponding to various damaged states of interest. The fact that such data are limited and mostly available from experimental campaigns, e.g. as in Galanopoulos et al. \citep{Galanopoulos2023}, has generated interest in population-based methods (see Tsialiamanis et al. \citep{Tsialiamanis2024}) that seek to transfer knowledge across similar structures in order to generalize prognostic models.

On the other hand, model-based methods for prognostics are not as strongly dependent on the availability of data obtained from deteriorated structures, which makes them more attractive for deterioration monitoring, and ultimately predictive maintenance, applications. However, this also comes with a caveat; namely, they require prior knowledge on the type of deterioration acting on the structure as well as its evolution through time. This prior knowledge is typically expressed in the form of assumed deterioration models that feature some degree of engineering or physical knowledge and are stochastic in nature \cite{Jia2018, Straub2020}. A typical example of this would be a model describing crack growth under fatigue loading. Furthermore, they include an additional level of prior knowledge, which is encoded in the model employed to simulate the behavior of the structural, or more generally, engineered system. In this work, this will be referred to as the observation model, drawing on the fact that deteriorating structural systems can been viewed as a class of dynamical systems \citep{Kamariotis2023, Cristiani2021}. For the crack propagation example, this observation model could be a FE model which requires information on both the length and orientation of the crack to serve its purpose.

As such, deterioration and/or observation model selection arises as a critical task in designing reliable SHM systems, since it singularly affects the quality of risk-informed decision support for O\&M that they seek to offer. As a problem, it is characterized by significant uncertainty, since the prior space over models can be wide-ranging in the sense that it may feature candidate models of varying degrees of fidelity and complexity. Furthermore, it is affected by the level of domain knowledge that is incorporated in both the deterioration process as well as the observation model. Uncertainty exists in both, through the parameters used to define the former as well as structural (e.g., material) or operational (e.g., loading) characteristics affecting the latter. Framing the problem in a Bayesian setting offers a natural means of reducing this uncertainty, while at the same time providing a set of tools that can be used to facilitate a quantitative assessment of different candidate models.

The advantages offered by the Bayesian approach were recognized by Beck et al. \citep{Beck2004}, who employed an asymptotic expansion of the Bayesian evidence to select the class of models that better describes the available data in a set of structural dynamics problems. This approach is consistent with the principle of the Bayesian Occam's razor (see e.g., McKay \citep{MacKay1992}), which naturally guards against the selection of overparameterized models that would tend towards overfitting and poor predictive performance. In a similar vein, Yin et al. \citep{Yin2019} employed the Bayesian evidence term to inform Finite Element (FE) model selection in dynamic reduction-based SHM. More recently, Koune et al. \citep{Koune2023} used it in a system identification context to assess different probabilistic models of the prediction error that account for potential temporal and spatial correlation. Estimating the Bayesian evidence term is characterized by significant computational complexity, due to it being the product of often high-dimensional intractable integrals. This has led to alternative approaches to model selection that employ likelihood-free or Approximate Bayesian Computation (ABC) algorithms which relax the need for explicit likelihood function formulations and have been successfully employed within the context of system identification in structural dynamics applications \citep{BenAbdessalem2019, Nayek2023}.

While model selection based on Bayesian evidence is a rigorous and elegant option, it provides an assessment of the candidate model class only with respect to its ability to recover quantities of interest (QoIs) based on the available data. It cannot assess the quality of the estimated QoIs with respect to downstream tasks related to predictive maintenance. It does not account for cost or quantify the confidence of the system designer in their modeling choice. Therefore, it does not allow for a notion of inductive bias to be introduced into the space of models. Ultimately, appropriate metrics are required to quantify the quality of inference conditioned on a specific model, which incorporate these considerations. Having identified a research gap in this area, the authors aim to propose a comprehensive framework to addess this issue.

This will be achieved through the use of tools from Bayesian decision theory \citep{Berger1985}, which casts the model selection problem as one of decision making under uncertainty (DMUU) \citep{kamariotis2023bms}.
It should be noted that decision-theoretic concepts are being used by the SHM community. Primarily it has been employed for SHM-aided maintenance planning either in the form of posterior decision analysis \citep{Vega2022, Chadha2023}, or in a pre-posterior analysis context where quantities such as the value of information (VoI) or value of SHM (VoSHM) are used to justify investment in an SHM system \citep{Kamariotis2023b} or for optimal sensor placement (OSP) \citep{Chinchilla2020}. Additionally, Hughes et al. \citep{Hughes2021, Hughes2022} have employed decision-theoretic principles to propose risk-based active learning, an online strategy where data labels in classification tasks are queried according to the decision 
support context in which these classifiers are employed. 

Ultimately, this work has employed ideas proposed in Kamariotis et al. \citep{kamariotis2023bms} and has aimed to extend them to propose a comprehensive model assessment framework used for diagnostics and prognostics in slowly evolving deterioration processes. This, to the authors best knowledge, is proposed for the first time. The effectiveness of the proposed framework is demonstrated for corrosion-induced thickness loss (CITL), which is a typical form of deterioration encountered in marine structures. A novel experimental programme has been designed and performed in lab-scale conditions that serves the purposes of the problem at hand.

The paper is laid out as follows. Section \ref{Seq2} presents the main theoretical underpinnings of the proposed framework, including general considerations on Bayesian model updating (BMU), as used in the context of SHM, and decision theory, as well as the proposed expected utility definitions and the motivation behind them. Section \ref{Seq3} deals with a description of the experimental case study and the dataset employed throughout this work. Section \ref{Seq4} demonstrates the implementation of the proposed framework across SHM tasks and Section \ref{Seq 5} offers concluding remarks. 

\section{Bayesian decision-theoretic model assessment} \label{Seq2}

In this section, we discuss the proposed decision-theoretic model assessment framework. The section begins by introducing key notions of Bayesian inference, in the form of BMU as typically used within SHM, as well as Bayesian decision theory. The latter is then placed within the context of model assessment and the key definitions of the proposed framework are provided.

\subsection{Bayesian model updating} \label{SubSeq2.1}

Let us begin by considering a generic BMU problem for SHM. We shall denote available observations from a monitored structure as a vector $\varepsilon \in \mathbb{R}^{N}$, where $N$ is the number of available observations; we employ this notation $( \varepsilon)$ for consistency purposes as the data used in this work correspond to mechanical strain. Let us also consider a vector of random variables (RVs) $\theta \in \mathbb{R}^{d}$, which ultimately constitute the QoIs in the BMU problem. These typically refer to unobserved quantities of the structural system that we seek to infer from data, such as material parameters that encode information on structural deterioration. In the case considered in this work, these may be observation model parameters or the parameters of a deterioration model that describes its evolution over time.

Using Bayes' theorem, we can obtain the ultimate goal of any Bayesian analysis, namely the posterior distribution over the parameters $\theta$ given a set of observations $\varepsilon$, as follows:
\begin{equation} \label{Eq1}
p ( \theta \vert \varepsilon) = \frac{p(\varepsilon \vert \theta) p( \theta)}{\int_{\mathcal{D_{\theta}}}p(\varepsilon \vert \theta) p( \theta) \, d\theta}
\end{equation}
where \( p(\varepsilon \vert \theta) \) denotes the likelihood function, \(p(\theta) \) is the prior over the QoIs and the integral in the denominator is the Bayesian evidence term. The choice of prior may be either empirical or subjective, but ultimately, rests on the analyst.
For this class of problems, constructing the likelihood function relies on an assumption on the probabilistic model that describes the discrepancy between the observations $\varepsilon$ and the predictions obtained from the observation model of the system, which is a function of the parameters.

Let us denote the model as $\mathcal{M}$ and its predictions as \( \hat{\varepsilon} =  \mathcal{M}(\theta)\); then, under the commonly employed uncorrelated Gaussian assumption for the prediction error \citep{Simoen2015}, we may write the likelihood function as:
\begin{equation} \label{Eq2}
p ( \varepsilon \vert \theta ) = \prod_{i=1}^{N}\mathcal{N}(\varepsilon_i - \hat{\varepsilon}_i;0 , \sigma^2)
\end{equation}
where in the above equation stochastic independence has been assumed for the observations, leading to the joint likelihood being expressed as the product of individuals. Often, the standard deviation term $\sigma$ is also inferred from data, and thus can be collected in the parameter vector.

The final ingredient in Eq. (\ref{Eq1}), i.e., the evidence term, is tractable only for specific prior-likelihood combinations, which are known as conjugate pairs \citep{Gelman2013}. In most cases of BMU for SHM, such pairs are not available; this is the case here as well due to our choice of priors. Instead, we have elected to adopt a sampling-based solution where inference is performed using Markov Chain Monte Carlo (MCMC) methods \citep{Brooks2011}; more specifically the no U-turn variant (NUTS) of Hamiltonian Monte Carlo (HMC) \citep{Hoffman2011, Betancourt2017}. Implementation is carried out in the Python probabilistic programming library Numpyro \citep{phan2019composable, bingham2019pyro}. For a more in depth exposition on HMC or NUTS, the reader is encouraged to study the provided references; further discussion will be avoided in the interest of concision. Additional, more detailed information on the Bayesian model structure will be provided in Section \ref{Seq4}, where implementation is discussed.

\subsection{Principles of Bayesian decision theory} \label{SubSeq2.2}

When viewing the task of assessing candidate models for structural deterioration monitoring as one of DMUU, the decision we are tasked with making revolves around the selection of the model which is more useful for a particular task. Bayesian decision theory provides a formal mathematical construct that enables the quantification of the cost associated with a particular decision. The notion of expected utility \citep{Barber2012} is central to the assessment process. Let us denote the decision to select a \textit{particular model} as $\alpha$; we can then use the output of a BMU analysis, namely the posterior distribution \( p(\theta \vert \varepsilon) \) of the parameters of that \textit{particular model} given the data, to define the expected utility as follows:
\begin{equation} \label{Eq3}
\mathcal{U}(\alpha) = \mathbb{E}_{p(\theta \vert \varepsilon)} \left[ U(\alpha, \theta) \right]
\end{equation}
where \( \mathbb{E}_{p(\theta \vert \varepsilon)} \) denotes the expected value operator with respect to the probability density function (pdf) of the outcome of decision $\alpha$, i.e., the BMU posterior. The function $U(\alpha, \theta)$ is known as the utility function and is nominally selected by the decision-maker; it assigns a utility value to a decision $\alpha$, typically ranging in $[0,1]$ with 1 expressing perfect utility, based on the set of outcomes $\theta$ that it has led to. A perfect utility implies that the decision made has led to the ideal outcomes which the decision-maker could expect; in the context of this work, it implies a model that perfectly explains the data and performs optimally in downstream tasks. In our case, for notational convenience the calligraphic $\mathcal{U}$ has been employed to denote the expected utility. For the purposes of model assessment, the set of possible decisions can be denoted as \( \alpha = \left\lbrace \mathcal{M}^{(j)} \right\rbrace_{j=1}^{M} \), where each member denotes a particular model choice indexed by $j$ and $M$ denotes the total number of available models. Moving forward, we shall use $\mathcal{M}^{(j)}$ instead of $\alpha$ to denote a particular decision for the sake of clarity.

Selecting the appropriate form for the utility function can be interpreted as a reflection of the decision-maker's risk profile, as discussed in Chadha et al. \citep{Chadha2023}. For the purposes of this work, a risk-neutral approach is followed which translates to a linear utility function. While assuming a risk profile provides added value for real-world structures managed under different circumstances, it was considered out of scope for the demonstrative case study employed here. Selecting an appropriate input for this function is crucial, as it must be such that it accurately reflects the performance of an action-outcome pair for a specific task, while also accounting for the potential effect of the cost associated with every action, which in our case represents modeling and computational cost.

\subsection{Defining expected utilities for model assessment} \label{SubSeq2.3}

What follows is an attempt by the authors to define such task-specific attributes and corresponding utility functions for structural deterioration monitoring applications and is the main contribution to the state of the art.
 
\subsubsection{A data-based view} \label{SubSubSeq2.3.1}

A natural choice for an attribute on which to define the utility of a model is its accuracy in terms of recovering the experimental observations, i.e., its ability to explain the data. A typical way to assess this would be to define the utility function based on a performance metric commonly used in regression tasks. One such choice is the Normalized Mean Square Error (NMSE), which can be defined as:
\begin{equation} \label{Eq4}
NMSE \left( \mathcal{M}^{(j)}, \theta_i \right) = \frac{100}{N \sigma^2_{\text{obs}}} \sqrt{\left( \varepsilon - \mathcal{M}^{(j)} \left( \theta_i \right) \right)^{\top} \left( \varepsilon - \mathcal{M}^{(j)} \left( \theta_i \right) \right)}
\end{equation}
where $\sigma^2_{\text{obs}}$ denotes the variance of the observations, $N$ is their number and $\theta_i$ refers to a particular posterior realization of the random vector of parameters. In this formulation, the NMSE is optimal when equal to 0 and equivalent to taking the observation mean when equal to 100. Therefore, the following NMSE-based utility function can be defined:
\begin{equation} \label{Eq5}
U \left( NMSE \left( \mathcal{M}^{(j)}, \theta_i \right) \right) = 1 - \frac{1}{100}NMSE \left( \mathcal{M}^{(j)}, \theta_i \right)
\end{equation}
and based on that the NMSE-based expected utility of a particular model choice is given by:
\begin{equation} \label{Eq6}
\mathcal{U}_{NMSE} \left( \mathcal{M}^{(j)} \right) = \mathbb{E}_{p(\theta \vert \varepsilon)} \left[ U \left( NMSE \left( \mathcal{M}^{(j)}, \theta_i \right) \right) \right] \approx \frac{1}{n_{\text{pos}}} \sum_{i=1}^{n_{\text{pos}}} U \left( NMSE \left( \mathcal{M}^{(j)}, \theta_i \right) \right)
\end{equation}

Based on its definition, the NMSE is incapable of providing a complete assessment of the model utility in cases where the prediction error standard deviation is included in the parameter random vector. Albeit not a parameter of the model itself, this QoI serves to quantify aleatoric uncertainty in the system. Therefore, it can provide an added benefit to model assessment through a utility function that penalizes models that compensate for poor predictive performance over the structural QoIs by inflating the estimated prediction error standard deviation. In this work, the authors indeed aim to provide such a definition. First, we begin by defining the relevant model attribute, which in this case is the likelihood of the posterior parameters conditional on the observed data. This is defined, in logarithmic form, for reasons of numerical stability, as follows:
\begin{equation} \label{Eq7}
\mathcal{L} \left( \mathcal{M}^{(j)}, \theta_i, \sigma_i \right) = \sum_{n=1}^{N} \left [ \log \mathcal{N}\left(\varepsilon_n - \mathcal{M}^{(j)}_{n}(\theta_n); 0, \sigma_{n}^2 \right) \right ]
\end{equation}
and the corresponding proposed utility function is defined as:
\begin{equation} \label{Eq8}
U \left( \mathcal{L} \left( \mathcal{M}^{(j)}, \theta_i, \sigma_i \right) \right) = 1 - \frac{\left \lvert \mathcal{L} \left( \mathcal{M}^{(j)}, \theta_i, \sigma_i \right) - \mathcal{L} \left( \mathcal{M}^{(j)}, \theta_{\text{MAP}}, \sigma_{\text{obs}} \right) \right \rvert}{\max \left \lbrace \left \lvert \mathcal{L}^{(j)}\left( \mathcal{M}^{(j)}, \theta_i, \sigma_i \right) \right \rvert, \left \vert \mathcal{L} \left( \mathcal{M}^{(j)}, \theta_{\text{MAP}}, \sigma_{\text{obs}} \right) \right \rvert \right \rbrace}
\end{equation}

We shall retain the dedicated notation $\sigma$ for the prediction error standard deviation and $\theta$ for the structural QoIs for clarity. The second term on the right hand side of Eq. (\ref{Eq8}) normalizes the log-likelihood, which spans the entire set of real numbers, to $[0,1]$ while ensuring that an optimal utility tends to 1. This is achieved through the use of the absolute value operator and by introducing the log-likelihood corresponding to the maximum a posteriori (MAP) estimate of the structural QoIs and the actual variance of the observations.

Under this definition, models are assigned a higher utility when they feature simultaneously narrow posteriors over the model parameters and estimates of the prediction error standard deviation that are similar to that exhibited by the data. The latter part serves as a guard against models that tend to inflate the prediction error variance to compensate for poorly inferred structural parameters. It should be noted that this holds under an assumption of homoscedastic noise. Finally, the maximum operation in the denominator ensures that values are constrained to the [0,1] range. Since the logarithm is a monotone increasing function, it ensures that for divergent log-likelihood terms in the numerator the utility function still tends to zero. The expected likelihood-based utility can be defined as follows:
\begin{equation} \label{Eq9}
\mathcal{U}_{\mathcal{L}} \left( \mathcal{M}^{(j)} \right) = \mathbb{E}_{p(\theta \vert \varepsilon)} \left[ U \left( \mathcal{L} \left( \mathcal{M}^{(j)}, \theta_i, \sigma_i \right) \right) \right] \approx \frac{1}{n_{\text{pos}}} \sum_{i=1}^{n_{\text{pos}}} U \left( \mathcal{L} \left( \mathcal{M}^{(j)}, \theta_i, \sigma_i \right) \right)
\end{equation}

\subsubsection{A decision support-based view} \label{SubSubSeq2.3.2}

Being able to effectively explain the data is not a guarantee of a model being useful for downstream SHM tasks. To illustrate this point, it is useful to consider that members of the model class, i.e., $ \mathcal{M} \in \mathcal{D}_{\mathcal{M}}$, are expected to differ in several ways. Namely, there may be analytical or numerical models, the latter potentially built using different discretization schemes or methods. For example, the stress intensity factor for a propagating crack can be calculated using analytical formulas or through FE analysis. Inevitably, each of them will contain some degree of domain knowledge about the type of structural deterioration form, since it is necessary to parameterize them with respect to some QoI describing it; otherwise, they would be unfit for model-based SHM. It is to be expected that different models will feature different parameterizations, either due to different levels of available domain knowledge or by virtue of their mathematical construction, One such example is the parameterization with respect to thickness in FE models of thin-walled structures that use 2D plate or 3D shell elements.

This difference may not affect their ability to explain the data, and therefore undermine their corresponding utility, but could affect their utility in light of decision support-related tasks, such as accurately estimating the probability of exceeding a threshold stress value. In this vein, the authors have sought to propose an additional definition of utility based on the ability of the updated model to accurately estimate the probability of exccedance of a particular limit state, defined in this context as:
\begin{equation} \label{Eq10}
P_{\text{f}} \left[ \mathcal{M}^{(j)}, \theta_i \right] =  \mathbb{P} \left [ g(x, \theta_i) = r(x) - s(\theta_i) < 0 \right ] =  \int_{- \infty}^{0}p_g(x, \theta_i) \, dx
\end{equation}
where $r(x)$ denotes the system capacity and $s(\theta_i)$ denotes the system demand for the $i$-th posterior realization. The limit state function $g$ is defined such that a failure event occurs when demand exceeds capacity and is described by the pdf $p_g$, obtained via the transformation of RVs.

In the context of this work, demand will be translated to the maximum von Mises stress $\sigma^{\max}_{\text{vm}}(\theta_i)$, calculated using the candidate model $\mathcal{M}$ for the $i$-th posterior realization. Capacity on the other hand will be described using the yield stress $\sigma_{\text{y}}$, with its pdf $p(\sigma_{\text{y}})$ considered as known. This corresponds to a typical structural reliability problem with a limit state based on elastic yielding. However, the definition is general and can be adapted based on the application for which the model is employed. Clearly, models of differerent fidelity, and potentially different parametric structure, are expected to yield different estimates of that quantity; yet, it is a quantity against which all models can be compared on equal grounds.

However, the question remains as to how to assign utility to the decision to select a particular model using the resultant probability of exceedance estimate. To address this, we propose the notion of an \textit{oracle} model $\mathcal{M}^{(o)}$; namely, a high-fidelity model built using \textit{a posteriori} information about the deteriorated structure and without any cost-related considerations. This represents an advanced level of domain knowledge and constitutes a singular modeling choice, which is meant to serve as a guide towards the selection of lower fidelity modeling options. Thus, we define a decision support-based utility function as:
\begin{equation} \label{Eq11}
U \left( P_{\text{f}} \left[ \mathcal{M}^{(j)}, \theta_i \right] \right) = 1 - \frac{\left \lvert \log_{10} P_{\text{f}} \left[ \mathcal{M}^{(o)}, \theta_i \right] - \log_{10} P_{\text{f}} \left[ \mathcal{M}^{(j)}, \theta_i \right] \right \rvert}{\max \left \lbrace \left \vert \log_{10} P_{\text{f}} \left[ \mathcal{M}^{(o)}, \theta_i \right] \right \rvert, \left \lvert \log_{10} P_{\text{f}} \left[ \mathcal{M}^{(j)}, \theta_i \right] \right \rvert \right \rbrace}
\end{equation}
where the logarithm base 10 is employed to alleviate numerical instability related to small numbers associated with the probabilities and the same approach to normalization is used as in the likelihood-based definition. Ultimately, the expected utility can be defined as:
\begin{equation} \label{Eq12}
\mathcal{U}_{P_{\text{f}}} \left( \mathcal{M}^{(j)} \right) = \mathbb{E}_{p(\theta \vert \varepsilon)} \left[ U \left( P_{\text{f}} \left[ \mathcal{M}^{(j)}, \theta_i \right] \right) \right] \approx \frac{1}{n_{\text{pos}}} \sum_{i=1}^{n_{\text{pos}}} U \left( P_{\text{f}} \left[ \mathcal{M}^{(j)}, \theta_i \right] \right)
\end{equation}

\subsubsection{A unified definition} \label{SubSubSeq2.3.3}

The availability of sufficient domain knowledge to construct an oracle model is by no means guaranteed. It is therefore important to account for the level of confidence the potential SHM system designer has on available information from deteriorated structures. Thus, we propose a unified definition of the expected utility which is based on a weighted average of the two previously defined components, namely:
\begin{gather} \label{Eq13}
\mathcal{U} \left( \mathcal{M}^{(j)} \right) = w_1 \cdot \mathcal{U}_{d} \left( \mathcal{M}^{(j)} \right) + w_2 \cdot \mathcal{U}_{P_{\text{f}}} \left( \mathcal{M}^{(j)} \right) \\
w_1,w_2 \in [0,1], \quad w_1 + w_2 = 1
\end{gather}
where $d$ denotes the data-based utility which could use either the NMSE, or log-likelihood attribute. This definition ensures that the utility function is bound in the $[0,1]$ range; it also allows for each individual definition to be retrieved by simply zeroing out the corresponding weight. By setting $w_2 = 0$, one retrieves the case where no information exists based on which an oracle model can be constructed. One should interpret the weights as reflecting the confidence of the analyst in the employed models and also the level of information that is available during system design. As an example, an analyst with significant confidence on the available levels of domain knowledge, and therefore model specification, may choose to place greater weight on the decision support-related task. Ultimately, the importance of domain knowledge regarding the deterioration form is paramount as will be shown later on.

\section{Experimental case study} \label{Seq3}

This section presents the experimental programme that was performed in order to demonstrate the proposed framework.

\subsection{Accelerated corrosion under three-point bending experiment} \label{SubSeq3.1}

Most structural systems currently in operation are exposed to some degree of corrosion deterioration; this is especially true for marine structures, which is the primary field of the authors. Despite recent interest in deploying SHM systems for CITL monitoring \citep{Kamariotis2023,Ghasemzadeh2023,Katsoudas2023,Silionis2024}, the naturally slow evolution of the phenomenon has constrained investigations to a numerical setting. In response to this, a novel testing procedure has been employed to achieve substantial CITL in a laboratory setting in a significantly accelerated timeframe, that is presented here for the first time.

\begin{figure}[!t]
	\centering
	\includegraphics[scale=0.95]{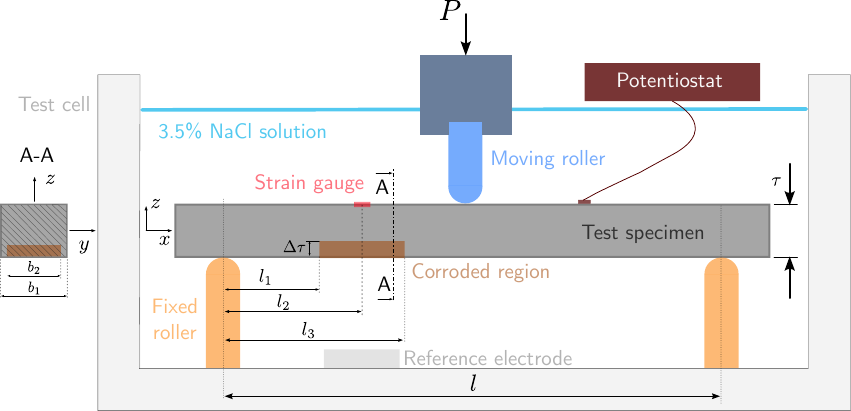}
	\caption{Schematic representation of accelerated corrosion under three-point bending test set-up. Details: $l = 77.86 \ \mathrm{mm}$, $l_1 = 21.27 \ \mathrm{mm}$, $l_2 = 26.54 \ \mathrm{mm}$, $l_3 = 32.24 \ \mathrm{mm}$, $b_1 = 29.32 \ \mathrm{mm}$, $b_2 = 25.56 \ \mathrm{mm}$, $\tau = 5.9 \ \mathrm{mm}$, $P = -1 \ \mathrm{kN}$.}
	\label{Fig 1}
	
\end{figure}

The experimental set-up, shown in Figure \ref{Fig 1}, consists of a rectangular prismatic steel specimen which is placed inside a watertight container, the test cell. The specimen is submerged in a corrosive environment, namely a solution consistent with saltwater and loaded under three point bending conditions. Its surface is insulated apart from a single rectangular region which is left exposed (corroded region). The test cell is configured so as to facilitate an electrochemical reaction, in which the test specimen acts as the working electrode (anode) and therefore suffers material loss at an accelerated pace. This is achieved by soldering an electrical connection to the specimen, which in turn is connected to a potenstiostat, an instrument that enables an electric potential to be applied between the working electrode and a reference electrode \citep{Colburn2021}. The latter consists of a Calomel reference electrode connected to a platinum mesh placed opposite the exposed region to accelerate the reaction, and therefore increase CITL in the exposed region over a given timeframe.

\begin{figure}[!htb]
	\centering
	\includegraphics[scale=0.85]{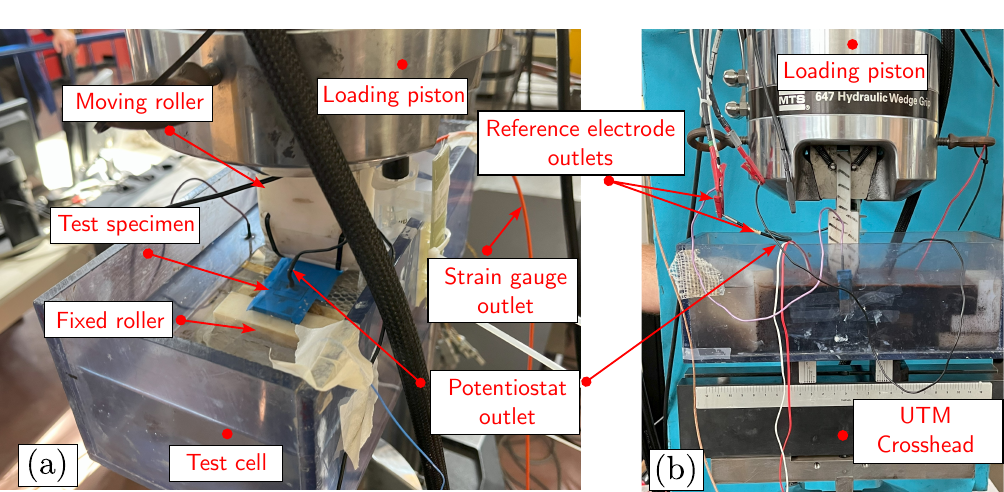}
	\caption{Laboratory implementation of the accelerated corrosion under three-point bending test in (a) perspective and (b) front view.}
	\label{Fig 2}
	
\end{figure}

At the same time as the reaction takes place, the specimen undergoes three-point bending. More specifically, it is placed on two fixed rollers, constructed using a non-reactive polymer, which are adhered to the test cell, which is itself securely connected to the fixed crosshead of a universal testing machine. A moving roller is attached to the loading piston of the machine to apply a constant amplitude force on the specimen, which is maintained throughout the duration of the test. A single resistance strain gauge measuring longitudinal strain is placed along the centerline of the specimen above the exposed region. Figure \ref{Fig 2} provides a snapshot from the laboratory implementation of the test.

\subsection{Data description} \label{SubSeq3.2}

The strain time series employed for the purposes of this work was obtained from a single test carried out over approximately 20 hours. The potentiostat was set to maintain a constant electric potential of 1.2 V on the working electrode; this value was selected to avoid overloading the circuit and therefore disrupting the test. Maintaining a stable potential is key to accelerating corrosion; if the electrical current is disrupted, then the specimen would undergo corrosion at a natural rate which would undermine the purpose of the test.

\begin{figure}[!b]
	\centering
	\includegraphics[scale=0.85]{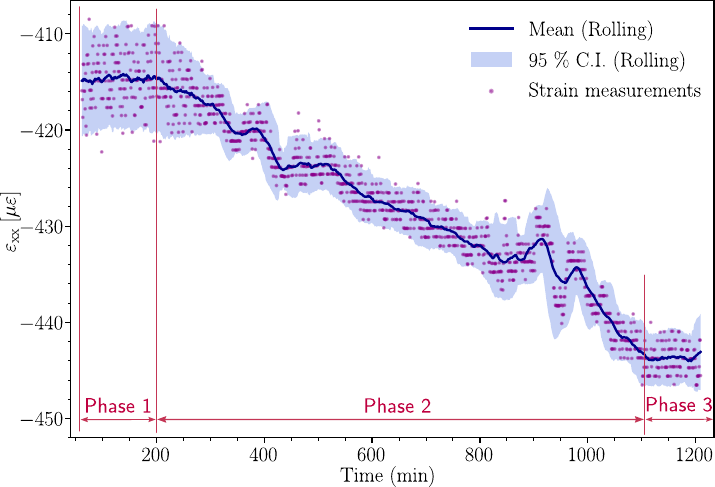}
	\caption{Strain measurements from the accelerated corrosion under three-point bending test.}
	\label{Fig 3}
	
\end{figure}

The strain data obtained during the test are plotted as a function of time in Figure \ref{Fig 3}; the scatter points correspond to the observations while a rolling mean curve and 95 \% confidence interval (C.I.) are also included. Until approx. 200 minutes into the test, a relatively constant mean strain can be observed with a homoscedastic variance structure; this is denoted as Phase 1. During that phase, it is apparent that CITL levels are such that no effective stiffness reduction has taken place. Thus, one may conclude that the specimen is still in its intact state and as such, the Phase 1 data will be employed for the distinct BMU task of inferring the Young's modulus of the specimen material. This was considered to be mild steel, but since in a real-world structure it is expected to exhibit variability, it was decided to include a system identification task to the demonstration for completeness.

After Phase 1, a decreasing trend is observed signifying an increase in the absolute strain magnitude, and therefore that CITL has begun in earnest. The relatively constant slope is expected since the constant voltage maintained by the potentiostat should lead to a relatively constant corrosion rate. This behavior, referred to as Phase 2, lasts up to 1100 minutes, and exhibits a slight heteroscedasticity, however at lower variance levels compared to Phase 1. At the tail of the time series, during Phase 3, the data revert to a similar behavior as in Phase 1, indicating a halt in corrosive activity; this can be attributed to the accumulation of corrosion byproducts, i.e., material waste, which is consistent with visual observations and is a feature of naturally occurring CITL.  Phase 3 data will also be  used for a dedicated task, namely Bayesian CITL identification, which in this context refers to using thickness loss directly as a QoI in the Bayesian setting.

\begin{figure}[!htb]
	\centering
	\includegraphics[scale=0.70]{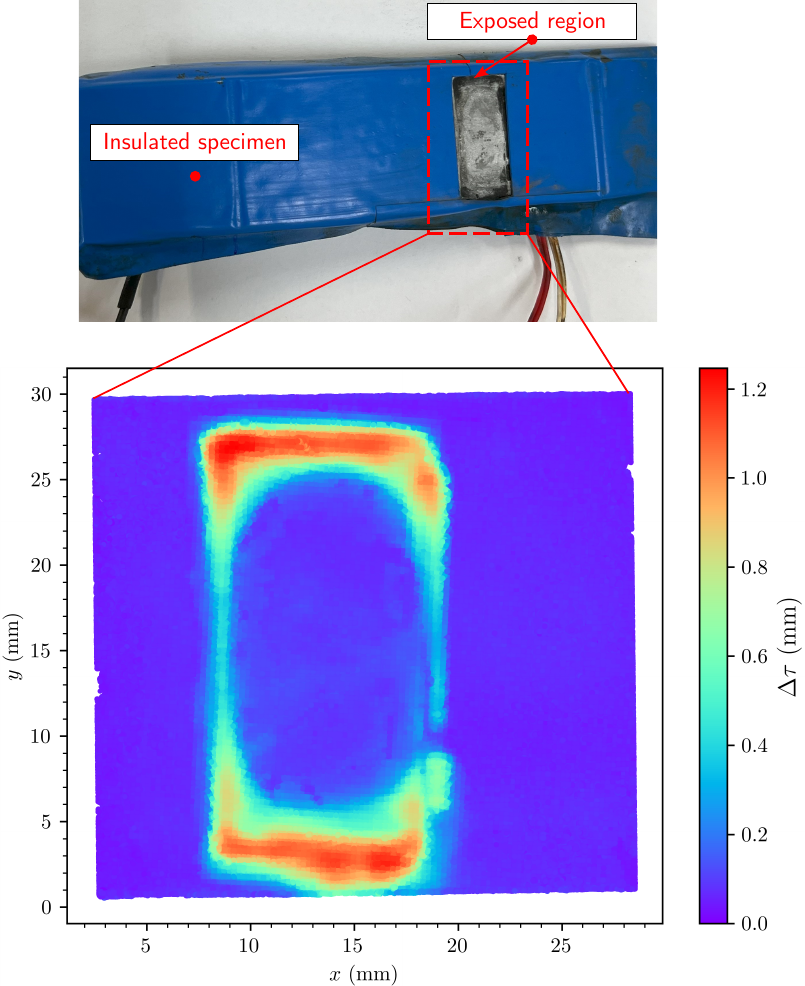}
	\caption{Close-up view of the specimen after the test (top) and heatmap of the 3D-scanned corroded surface (bottom). Axes notation follows Figure \ref{Fig 1}.}
	\label{Fig 4}
	
\end{figure}

After the conclusion of the test, a surface scan of the corroded region took place to assess the level of thickness loss that was achieved; the results of the scan are plotted in the form of a heatmap in Figure \ref{Fig 4}. It should be noted that the scale of the horizontal axis is expressed in a local coordinate system based on the scanning instrument. Nevertheless, it is evident that corrosion was not uniform over the exposed region; rather, more significant thickness loss can be observed along its periphery with the strips parallel to the edges of the specimen being more significantly affected.

\section{Model assessment across SHM tasks} \label{Seq4}

This section contains the results from the implementation of the proposed framework across different SHM tasks using the data obtained from the previously described experiment. Initially, the pool of candidate models will be described followed by implementation on three distinct tasks: inferring the Young's modulus of the material; inferring CITL directly; and finally, inferring the parameters of a CITL deterioration model for prognostics.

\subsection{Model specification} \label{SubSeq4.1}

We begin by specifying the pool of candidate models $\left \lbrace \mathcal{M}^{(j)} \right \rbrace _{j=1}^{M}$, which refers to the observation models used to construct the likelihood function in the Bayesian model. These have been selected so as to represent different levels of prior domain knowledge on the deterioration form; namely, the extent of the corroded area as well as the type of corrosion.

The first model, denoted as $\mathcal{M}^{(1)}$, corresponds to the lowest level of prior domain knowledge; it is an analytical model, based on Euler-Bernoulli beam theory, which considers uniform thickness loss across the entire specimen. Clearly, this model cannot accurately represent the localized stiffness reduction caused by the limited extent of the actual corroded region, as shown in Figure \ref{Fig 4}. Still, it is representative of a modeling assumption where the analyst is either incapable or unwilling to allocate more significant resources; needless to say, this model is associated with the lowest cost, as the following formula can be used to transform thickness loss to strain:
\begin{equation}\label{eb_form}
\varepsilon_{\text{xx}} = \mathcal{M}^{(1)}(E, \Delta \tau) = \frac{3Pl_2}{(\tau - \Delta \tau)^2b_1E}
\end{equation}
where the quantities contained therein are as reported in Figure \ref{Fig 1} and the parameterization is consistent with the tasks for which the model is to be employed later.

\begin{figure}[!htb]
	\centering
	\includegraphics[scale=0.85]{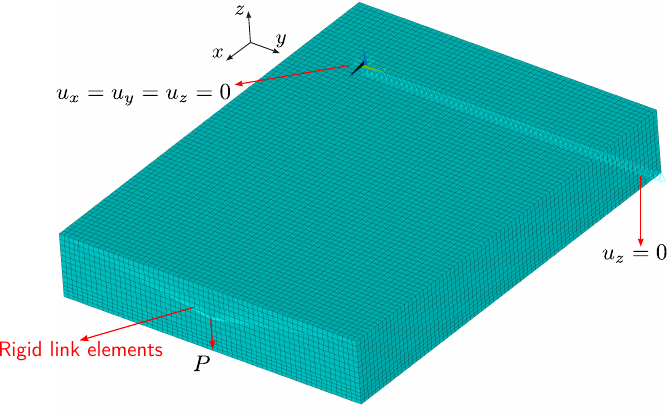}
	\caption{Boundary conditions and kinematic constraints for load application in FE models.}
	\label{Fig 5}
	
\end{figure}

For the second level of models we shall adopt a higher fidelity option, namely FE modeling. This will allow for an explicit modeling of the localized stiffness reduction, through a more detailed representation of the corroded region. More specifically, uniform thickness loss will be applied throughout the extent of the rust-colored area denoted in Figure \ref{Fig 1}, with the thickness loss $\Delta \tau$ serving as the model parameter. The geometry will be modeled in two different ways, dictated by the use of two element types, namely three-dimensional solid and shell elements.
Linear elements have been employed for all FE models; kinematic constraints were used to apply the loading action through a master-slave connection where rigid link elements connected the master to all nodes along the contact line of the moving roller. Boundary conditions along the contact lines of the fixed rollers were applied by simply constraining the appropriate degrees of freedom (DOFs). These are shown, indicatively for the solid element case, in Figure \ref{Fig 5} for one symmetric half of the model.

\begin{figure}[!htb]
	\centering
	\includegraphics[scale=0.85]{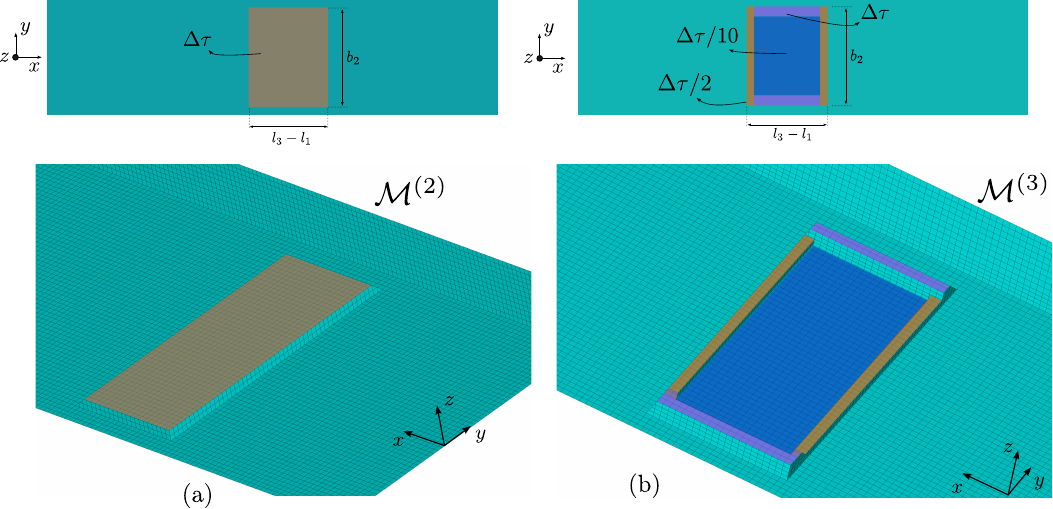}
	\caption{Close-up view of corroded region FE modeling approaches (a) without and (b) with \textit{a posteriori} information on the deterioration form. Axes notation follows Figure \ref{Fig 1}.}
	\label{Fig 6}
	
\end{figure}

In the solid element case, a three-dimensional representation of the geometry is required and the corresponding model is denoted as $\mathcal{M}^{(2)}$; a close-up view of it is shown in Figure \ref{Fig 6} (a). For the shell model, a midsurface model is constructed and the thickness loss is introduced through reassignment of cross-sectional properties for those elements located in the corroded region. Thus, the shell element model $\mathcal{M}^{(4)}$ contains a significantly lower number of total nodes for a similar element size and as a result requires a tenth of the (wall) time for solution compared to $\mathcal{M}^{(2)}$. However, one can expect $\mathcal{M}^{(2)}$ to be more accurate in terms of modeling localized effects on the resultant stress field due to its three-dimensional formulation and corresponding higher refinement. 

The second level of models corresponds to a higher level of prior domain knowledge on the deterioration form; however, it is still within reason in terms of being assumed \textit{a priori}. Thus, at the third level of available models we shall consider those containing \textit{a posteriori} information on the deterioration form, obtained from the 3d scanned surface (see Figure \ref{Fig 3}), and choose among them the oracle model.
Due to the complexity of the corroded surface, certain assumptions are required at this stage as well in order to construct FE models that strike a reasonable balance between accurate representation of damage and cost. As such, we have employed a modeling strategy whereby the corroded region is modeled so that three distinct zones are considered, each exhibiting uniform corrosion. This is illustrated in Figure \ref{Fig 6} (b) which shows a close-up view from the corresponding solid element model $\left( \mathcal{M}^{(3)} \right)$.

The parametric structure of this model, as well as its shell element counterpart $\left( \mathcal{M}^{(5)} \right)$, is centered around thickness loss $\Delta \tau$ in the strips parallel to the specimen edges, with thickness in the other zones expressed as a fraction of that quantity.
Namely, the strips parallel to the breadth of the specimen were assumed to exhibit $\Delta \tau / 2$, while the central area $\Delta \tau / 10$. This was based on observation of the final corroded surface morphology (see Figure \ref{Fig 4}) alongside the fact that the entire test is considered to occur under a constant corrosion rate. Among the two models based on \textit{a posteriori} domain knowledge, it has been decided to employ $\mathcal{M}^{(3)}$ as the oracle, due to the fact that it is expected to showcase incrased accuracy in representing the localized stress state in the vicinity of the corroded region.

As mentioned earlier, the shell element-based models require approximately 10\% of the wall time required by their solid element-based counterparts. Despite the fact this cost is not considerable in absolute terms, being on the order of 1 and 10 seconds per solution respectively, it still is significant when cast in a stochastic simulation setting. The proposed framework features both the implementation of MCMC methods, as well as a crude Monte Carlo (MC) estimator for the expected utility which requires repeated calls to the FE solver in the case of the decision support-based definition. As a result, it has been decided to use surrogate models in lieu of the original FE-based ones; to avoid confusion we will retain the same notation for the surrogates as for the original models on which they were trained.

\begin{figure}[!htb]
	\centering
	\includegraphics[scale=0.9]{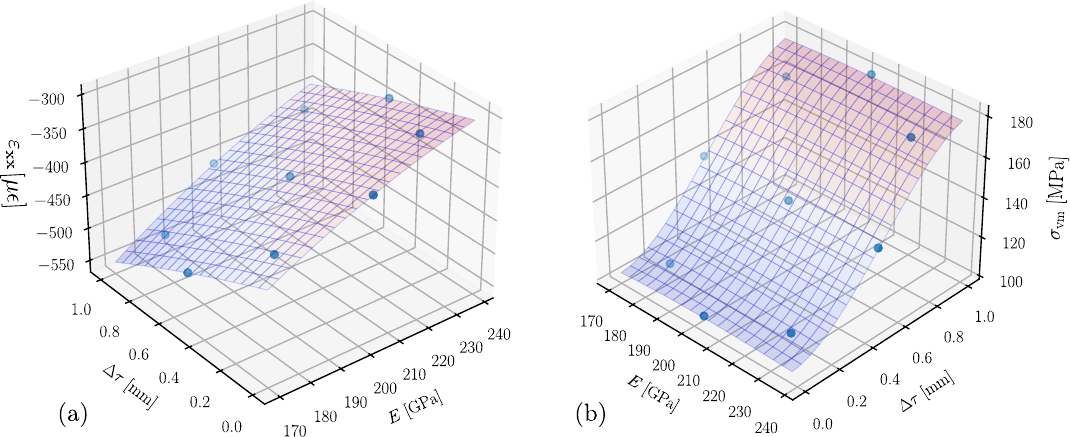}
	\caption{Polynomial regression surrogates of $\mathcal{M}^{(2)}$ for (a) strain and (b) maximum von Mises stress prediction.}
	\label{Fig 7}
\end{figure}

As the problem mechanics were relatively simple, it was decided to employ typical polynomial regression models. Alongside thickness loss $\Delta \tau$, the surrogate model input also included the Young's modulus $E$, to enable it to be used for material parameter identification as well. In total, two surrogates were trained for each candidate FE model; one to return strain $\varepsilon_{\text{xx}}$ and the other the maximum von Mises stress $\sigma_{\text{vm}}$ used to define system demand $s$. Their training inputs were obtained using Design of Experiments (DoE) over ranges of the input parameters defined based on the employed priors for the Young's modulus, while for thickness loss engineering judgement was employed for models $\mathcal{M}^{(2,4)}$ and information from the corroded surface for models $\mathcal{M}^{(3,5)}$. Table \ref{Tab2} contains details on the training inputs, the order of the polynomials used, as well as their training performance expressed through the $R^2$ metric. Representative polynomial response surfaces obtained for $\mathcal{M}^{(2)}$ for both surrogate tasks are shown in Figure \ref{Fig 7}.
	
\begin{table}[!htb]
  \centering
  \caption{Surrogate model training and performance details}
  \vspace{0.1cm}
    \begin{tabular}{ccccccc}
    \toprule
    Model & \multicolumn{2}{c}{Parameter ranges} & \multicolumn{2}{c}{Polynomial order} & \multicolumn{2}{c}{$R^2$} \\
    \midrule
          & $E$ [GPa] & $\Delta \tau$ [mm] & $\varepsilon_{\text{xx}}$ & $\sigma_{\text{vm}}$    & $\varepsilon_{\text{xx}}$ & $\sigma_{\text{vm}}$ \\
          \midrule
    $\mathcal{M}^{(2)}$    & [170.0, 238.0] & [0.0, 1.0] & 2     & 3     & 1.00     & 1.00 \\
   $\mathcal{M}^{(3)}$    & [170.0, 238.0] & [0.0, 1.4] & 2     & 3     & 1.00     & 1.00 \\
    $\mathcal{M}^{(4)}$    & [170.0, 238.0] & [0.0, 1.0] & 2     & 3     & 1.00     & 1.00 \\
    $\mathcal{M}^{(5)}$    & [170.0, 238.0] & [0.0, 1.4] & 2     & 3     & 1.00     & 0.96 \\
    \bottomrule
    \end{tabular}%
  \label{Tab2}%
\end{table}%

The use of polynomial regression models provides an added benefit with respect to the implementation of HMC methods, which was exploited in this work. More specifically, it enables the use of automatic differentiation \citep{Baydin2017} to estimate the derivatives required by HMC, thus negating the need for explicit derivation and providing significant acceleration in computation. This functionality is offered by Numpyro and JAX, which were used to write the code that implements the proposed framework. This allowed the authors to employ a state-of-the-art algorithm like NUTS and take advantage of its superior performance using relatively modest computational resources, namely a machine with 32 GB of RAM and an Intel\textsuperscript{®} Core\textsuperscript{TM} i7-10750H CPU.

\subsection{System identification: material parameter inference} \label{SubSeq4.2}

\begin{figure}[!t]
	\centering
	\includegraphics[scale=0.7]{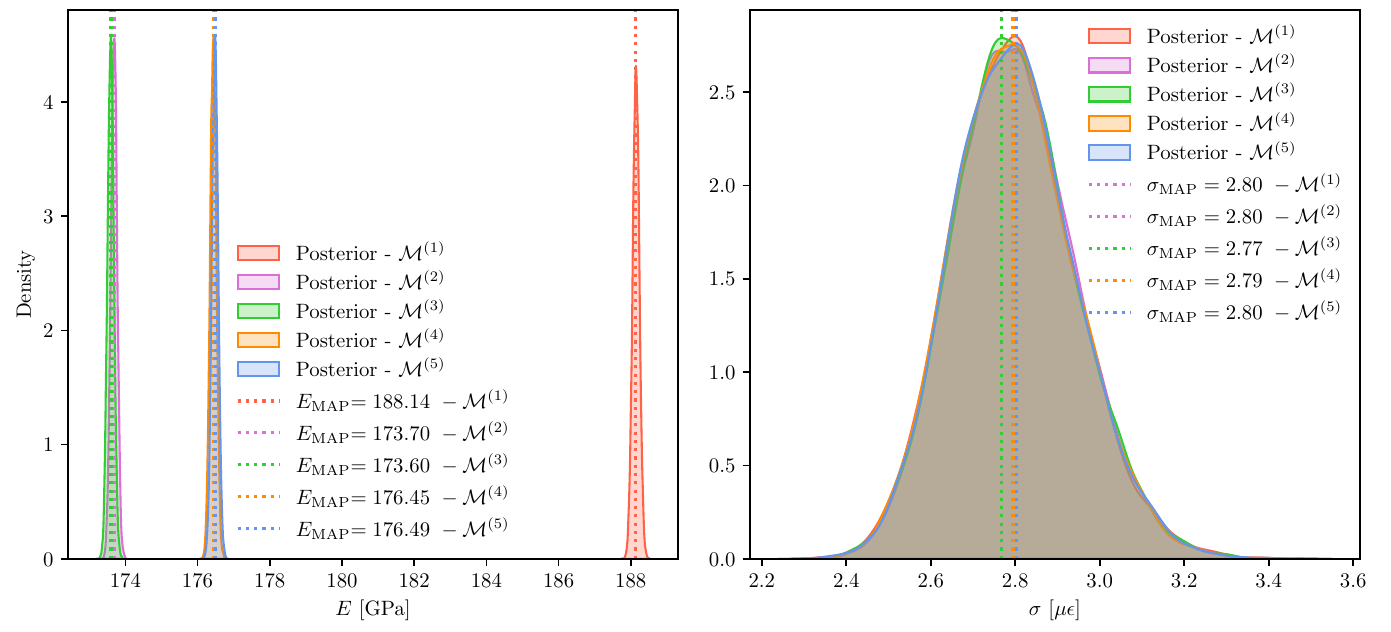}
	\caption{KDE-based posteriors of the Young's modulus (left) and the prediction error standard deviation (right).}
	\label{Fig 8}
	
\end{figure}

We shall first implement the proposed framework on a typical BMU task, namely inferring the Young's modulus $E$ using strain response data; the prediction error standard deviation will also be included as a QoI, thus leading to the parameter vector $\theta = \left \lbrace E, \sigma \right \rbrace$. The parameters were assumed to be independent, thus leading to a joint prior $p(\theta) = p(E)p(\sigma)$, where \( E \sim \mathcal{U}(170.0, 238.0) \ \mathrm{GPa}\) and \( \sigma \sim \mathcal{U}(0.0, 10.0) \ \upmu \upepsilon \). The range of the uniform prior for $E$ is consistent with the surrogate model input space. As is typical in MCMC, we derive the target log-posterior, where the logarithm is taken for computational purposes, through the numerator of Bayes' theorem, as:
\begin{equation} \label{Eq15}
\log p(\theta \vert \varepsilon) \propto \sum_{i=1}^{N} \left[ \log \mathcal{N} \left( \varepsilon_i - \mathcal{M}^{(j)}_{i}(E, \Delta \tau = 0); 0, \sigma^2 \right) \right] + \log p(\theta)
\end{equation}

where the input structure of the model has been explicitly defined for this task by considering $\Delta \tau = 0$. To sample from this posterior, we ran NUTS over 8 chains simultaneously with each one starting from a different seed; 2000 warm-up steps were used to initially tune the algorithm hyperparameters and find the typical set and an equal number of samples was drawn from the posterior for each chain. By running multiple chains we sought to ensure sufficient mixing, and therefore be confident that we are sampling from their stationary distribution, i.e., the true posterior.
As a quantitative means to monitor convergence we have employed the rank-normalized $\hat{R}$ diagnostic \citep{Vehtari2021} and accepted posteriors where it was below 1.01. This indicates that the between-chain and within-chain variances are near identical, thus justifying convergence. Figure \ref{Fig 8} plots the resultant posterior distributions using kernel density estimation (KDE) for both parameters using posterior samples from a single, converged, chain.

In terms of the Young's modulus, shown in the left panel, one observes very narrow posterior estimates, which is to be expected, alongside a clustering according to model definition; namely FE models cluster based on element type, and therefore discretization scheme, while the analytical model stands apart. This is a reflection of the expected discrepancy between different models used to explain the same physical problem, as well as the fact that they are equally capable in explaining the aleatoric uncertainty in the system. This is indicated by the right panel, where the posteriors of the prediction error standard deviation overlap.

The similarity between candidate models is further demonstrated by the fact that they exhibit equal, very high, data-based expected utilities; namely, $\mathcal{U}_{NMSE} \approx 0.97$ and $\mathcal{U}_{\mathcal{L}} \approx 0.99$, estimated using the posterior sample again from a single chain. The latter being higher demonstrates the added value of including the inferred prediction error standard deviation to the utility definition. Ultimately, for the tasks to come the Young's modulus used for each candidate model will be assumed equal to the posterior MAP estimate obtained here, and shown in Figure \ref{Fig 8}. Were one to interpret utilities purely as a means for model selection, then the analytical model would be superior in this instance on account of its lower cost.

\subsection{Diagnostics: thickness loss monitoring} \label{SubSeq4.3}

The second task that will be considered belongs to the diagnostic aspect of SHM viewed from a model-based perspective. Namely, the goal will be to infer the thickness loss $\Delta \tau$ directly from the strain response data from Phase 3, where we consider that the phenomenon is no longer active. In this case as well, we will also seek to infer the prediction error standard deviation yielding the parameter vector $\theta = \left \lbrace \Delta \tau, \sigma \right \rbrace$. A similar, independent prior structure will be assumed, that is \( p(\theta) = p(\Delta \tau)p(\sigma) \), where \( \Delta \tau \sim \mathcal{U} \left( 0.0, \Delta \tau^{(j)} \right) \ \mathrm{mm}\) and \( \sigma \sim \mathcal{U}(0.0, 10.0) \ \upmu \upepsilon \). The upper limit $\Delta \tau^{(j)}$ of the thickness loss prior is equal to the corresponding term used to train each surrogate model, as reported in Table \ref{Tab2}. In this case, the target log-posterior is defined by parameterizing the models in terms of the thickness loss, as follows:
\begin{equation} \label{Eq16}
\log p(\theta \vert \varepsilon) \propto \sum_{i=1}^{N} \left[ \log \mathcal{N} \left( \varepsilon_i - \mathcal{M}^{(j)}_{i}\left(E = E^{(j)}_{\text{MAP}}, \Delta \tau\right); 0, \sigma^2 \right) \right] + \log p(\theta)
\end{equation}

In terms of implementing NUTS, exactly the same parameters were used as in Section \ref{SubSeq4.2} and similarly satisfactory results were returned in terms of convergence across all models. It is interesting to point out here that running NUTS for $\mathcal{M}^{(2)}$ over 8 chains including warm-up requires approx. 16.6 seconds in terms of wall time; compared to the fact that the corresponding FE model required approx. 10 seconds for one solution further demonstrates the usefulness of the polynomial-based surrogate.

\begin{figure}[!htb]
	\centering
	\includegraphics[scale=0.7]{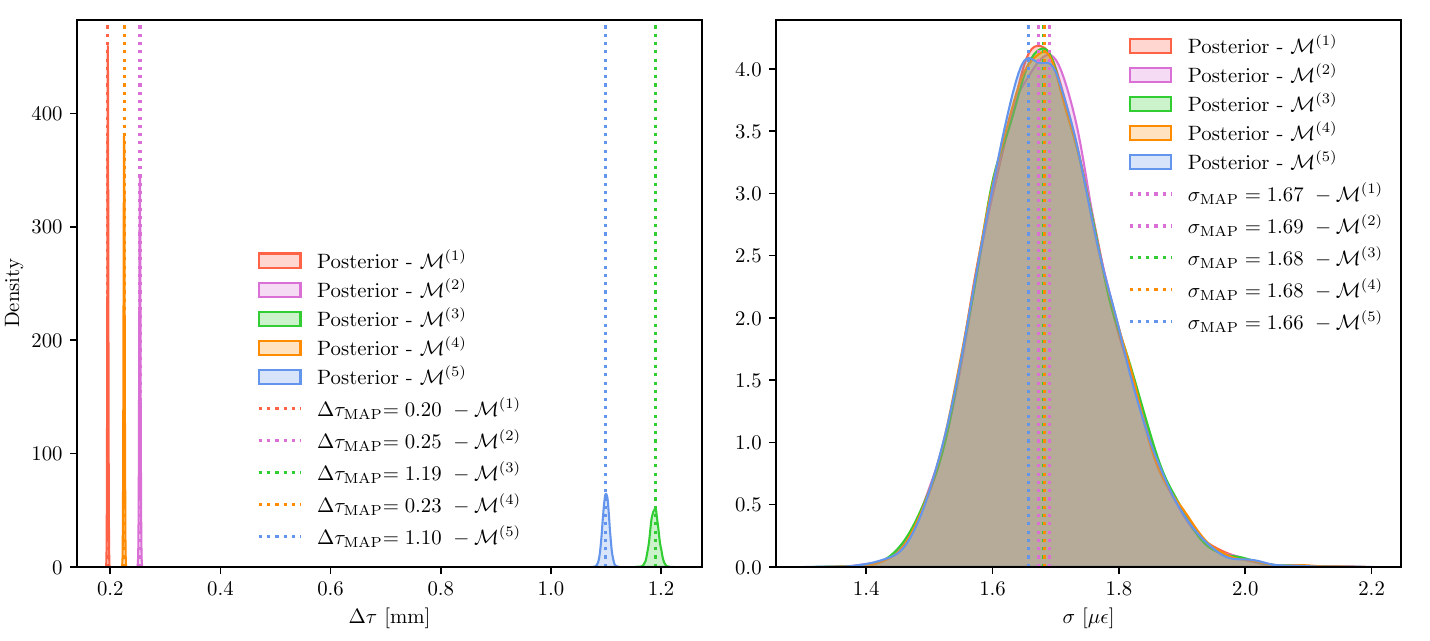}
	\caption{KDE-based posteriors of the thickness loss (left) and the prediction error standard deviation (right).}
	\label{Fig 9}
	
\end{figure}

In terms of the inference results, the KDE-based posteriors over the parameters are plotted in Figure \ref{Fig 9}; again, all candidate models are equivalent in describing the aleatoric uncertainty in the system as shown by their overlapping posteriors (right panel). The thickness loss posteriors exhibit very low uncertainty; as expected, a lower estimate is returned by $\mathcal{M}^{(1)}$, where uniform thickness loss was assumed for the entire specimen. Interestingly, the FE models featuring uniform thickness loss over the corroded region, i.e., $\mathcal{M}^{(2,4)}$, lead to thickness loss posteriors that exhibit greater similarity to that of the analytical model, than to those obtained using \textit{a posteriori} information.

\begin{figure}[!b]
	\centering
	\includegraphics[scale=0.9]{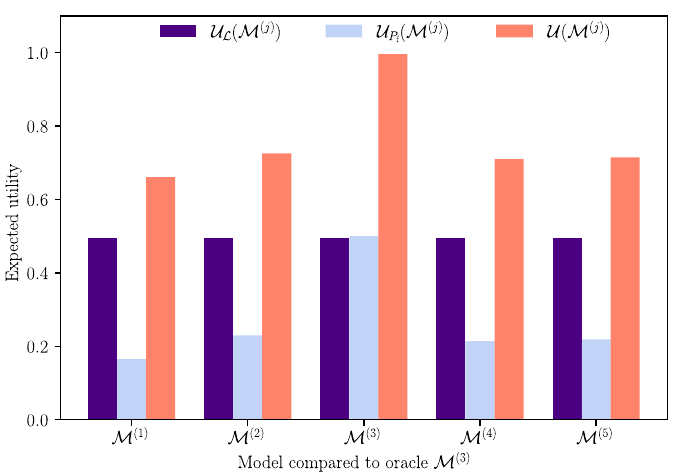}
	\caption{Expected utility of candidate models for the diagnostic SHM task considering $\mathcal{M}^{(3)}$ as the oracle.}
	\label{Fig 10}
	
\end{figure}

This provides a first indication into the importance of domain knowledge and the potential inability of a model selected based on an equivalent stiffness reduction-type reasoning to accurately capture a potentially complex deterioration form. However, the latter may not necessarily be an attribute that significantly influences decision making. Therefore, it would be interesting to view these results through the lens of the proposed unified definition of expected utility. This is demonstrated in Figure \ref{Fig 10}, where the unified expected utility is plotted in the form of a bar chart for the case of equally weighted components. For the probability of failure estimate, a Gaussian distribution was used for the yield stress, namely $\sigma_{\text{y}} \sim \mathcal{N}(284.5, 21.5)$ MPa \citep{Melcher2004}.

\begin{figure}[!htb]
	\centering
	\includegraphics[scale=0.9]{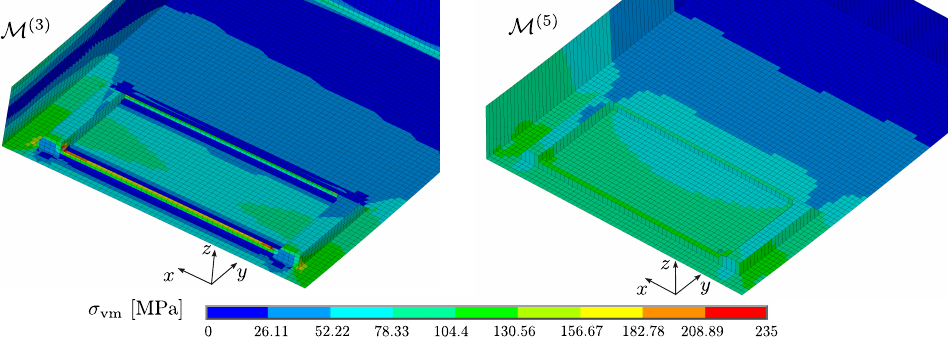}
	\caption{Contour plots of von Mises stress in the vicinity of the corroded region for models $\mathcal{M}^{(3)} \ \& \ \mathcal{M}^{(5)}$}
	\label{Fig 11}
	
\end{figure}

It becomes clear that according to the likelihood-based attribute, used in the data-based definition, the candidate models are equivalent and lead to near-perfect utility scores. However, their decision support-based utility, assessed according to their ability to estimate the failure probability compared to the oracle model, is considerably lower. The FE-based ones exhibit some improvement compared to the analytical model; however, their advantage is modest and under stricter cost considerations could favour a decision to select the analytical model. A further interesting conclusion may be gleaned by assessing the two shell element-based models, namely $\mathcal{M}^{(4)} \ \& \ \mathcal{M}^{(5)}$, which demonstrate virtually equivalent utility. In the case of the latter, high quality domain knowledge on the deterioration form is counterbalanced by a poor choice of element technology, as shells are proven unable to accurately capture the localized stress concentrations caused by the non-uniform thickness loss pattern of the actual specimen. This is further illustrated by Figure \ref{Fig 11} where contour plots of the von Mises stress are provided comparing shell and solid-based models $\mathcal{M}^{(3)} \ \& \ \mathcal{M}^{(5)}$.

\subsection{Prognostics: CITL forecasting} \label{SubSeq4.4}

The final task concerns monitoring structural deterioration in the form of CITL; this consists essentially of selecting a deterioration model that describes how this quantity evolves over time and framing inference around its parameters. In this work, we have elected to use two very commonly employed models found in the marine structures literature \citep{Paik2003, Garbatov2010}, which are defined as follows:
\begin{equation}\label{Eq17}
\Delta \tau_{\log}(t; \alpha, \beta, \gamma) = \frac{\gamma}{1 + \exp (- (\alpha + \beta t))} \\
\end{equation}
\begin{equation}\label{Eq18}
\Delta \tau_{\exp}(t; \alpha, \beta) = \frac{\alpha}{\Delta t_{\text{ts}}} \left( t-t_0 \right)^\beta
\end{equation}

The motivation behind using two alternative formulations for the deterioration model was to illustrate that the proposed framework is not restricted to the assessment of different observation models. The model in Eq. (\ref{Eq17}) is a three parameter logistic-type model which has commonly been employed to characterize CITL growth from historical data, obtained via on-site inspections from ocean going vessels. Its main advantage is the fact that it can model asymptotic behavior typically exhibited by naturally corroding structures; namely the fact that the phenomenon tends to halt after a period of time due to wastage accumulation, and thus CITL converges to a constant value. In a typical SHM context, this upper cut-off value, i.e., the numerator term $\gamma$, is not expected to be known, as opposed to modeling based on historical data where it may be available.
As a result, it will also be inferred from the data, leading to a vector of QoIs $\theta = \left \lbrace \alpha, \beta, \gamma, \sigma \right \rbrace$.
Again, we assume an independent joint prior with the following structure: 
\begin{gather} \label{Eq19}
\alpha \sim \text{Half-Normal}(0.1) \\
\beta \sim \mathcal{N}(0.0, 1.0) \\
\gamma \sim \mathcal{U} \left( 0, \Delta \tau^{(j)} \right) \\
\sigma \sim \mathcal{U}(0.0, 10.0)
\end{gather}
and the corresponding log-posterior can thus be stated as:
\begin{equation}\label{Eq20}
\log p(\theta \vert \varepsilon) \propto \sum_{i=1}^{N} \left[ \log \mathcal{N} \left( \varepsilon_i - \mathcal{M}^{(j)}_{i}\left(E = E^{(j)}_{\text{MAP}}, \Delta \tau_{\log} \left(t_i; \alpha, \beta, \gamma \right)\right); 0, \sigma^2 \right) \right] + \log p(\theta)
\end{equation}
where the QoIs are now introduced indirectly to the likelihood function through the deterioration model; $t_i$ refers to a discrete points in time when observations are available. Prior selection for $\alpha, \beta$ reflects general considerations about the shape of the curve, such as ensuring non-negativity, while the range where $\gamma$ is uniformly distributed is such that it reflects the upper training limits of the surrogates. This ensures that no out-of-range inputs appear during inference, which could jeopardize its quality.

The model of Eq. (\ref{Eq18}) is exponential in the parameters and has also found wide use in modeling historical CITL data. It is a typical two parameter model, featuring a rate term $\alpha$ and an exponent term $\beta$, and in the formulation adopted here the corrosion initiation time $t_0$ has been included, but considered deterministic. An additional normalizing constant, $\Delta t_{\text{ts}}$, has been included as well which equals the length of the timeseries over which data is available for inference and serves to downscale the rate term $\alpha$. This is required to avoid diverging chains triggered by the vastly different physical scale of this parameter compared to the others, which is unavoidable even when using a sophisticated algorithm such as NUTS. Ultimately, the parameter vector for this case becomes $\theta = \left \lbrace \alpha, \beta, \sigma \right \rbrace$ and we define the following structure for the independent priors:
\begin{gather} \label{Eq21}
\alpha \sim \mathcal{U}(0.1, 1.5) \\
\beta \sim \mathcal{U}(-1.5, 1.5) \\
\sigma \sim \mathcal{U}(0.0, 10.0)
\end{gather}
and the corresponding log-posterior:
\begin{equation}\label{Eq22}
\log p(\theta \vert \varepsilon) \propto \sum_{i=1}^{N} \left[ \log \mathcal{N} \left( \varepsilon_i - \mathcal{M}^{(j)}_{i}\left(E = E^{(j)}_{\text{MAP}}, \Delta \tau_{\exp} \left(t_i; \alpha, \beta \right)\right); 0, \sigma^2 \right) \right] + \log p(\theta)
\end{equation}
The priors adopted for the exponential-type deterioration model reflect our expectations of a deterioration model describing CITL. On the one hand, the rate term is strictly positive, which combined with the inclusion of the corrosion initiation time, ensures non-negativity. On the other hand, the range of the exponent term prior allows for both a concave and a convex shape, but restricts the order of the polynomial to avoid unreasonably rapid growth. NUTS was implemented using the same hyperparameters as mentioned in the other tasks; this time, however, a different treatment of the observations was considered.

\begin{figure}[htb!]
	\centering
	\includegraphics[scale=2.10]{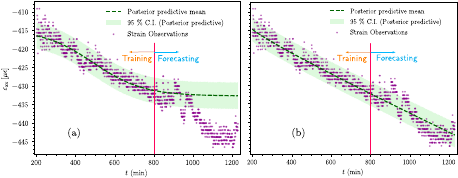}
	\caption{Posterior predictive processes using (a) the logistic-type and (b) exponential-type deterioration model. Results shown correspond to $\mathcal{M}^{(2)}$.}
	\label{Fig 12}
	
\end{figure}

To begin with, all strain data from Phases 2 \& 3 were considered; that is all measurements past minute 200 of the experiment. Subsequently, they were separated into two sets, one for training and one for forecasting purposes, namely $\varepsilon_{\text{tr}} = \left \lbrace \varepsilon \right \rbrace_{200 \leq t < 800}$ and $\varepsilon_{\text{fr}} = \left \lbrace \varepsilon \right \rbrace_{t \geq 800}$. The former, as the name suggests, was considered for typical inference purposes, i.e., it was used to obtain posteriors over the relevant parameter vector. The latter was employed to assess the forecasting qualities of the model by comparing it to its posterior predictive process, which in turn can be defined using the posterior parameters $\theta_{\text{pos}}$ inferred from the training set as:
\begin{equation}\label{Eq23}
p(\tilde{\varepsilon}_{\text{fr}} \vert \varepsilon_{\text{tr}}) = \int_{\mathcal{D}_{\theta_{\text{pos}}}} p\left(\tilde{\varepsilon}_{\text{fr}} \vert \theta_{\text{pos}}\right)p\left(\theta_{\text{pos}}\right) \, d\theta_{\text{pos}} 
\end{equation}
where \( p\left(\theta_{\text{pos}} \right) = p \left( \theta \vert \varepsilon_{\text{tr}} \right) \) is the posterior obtained using the training data. Essentially, the posterior predictive uses $ p\left(\theta_{\text{pos}} \right) $ to weigh the likelihood of strain realizations, obtained by propagating realizations from $ p\left(\theta_{\text{pos}} \right) $, to the deterioration and consequently observation model for the forecasting time horizon. The posterior predictive mean and 95 \% credible intervals are shown in Figure \ref{Fig 12} (a) for the logistic-type deterioration model and Figure \ref{Fig 12} (b) for the exponential-type one; both instances correspond to candidate model $\mathcal{M}^{(2)}$ and have been estimated from 1000 samples generated from the posterior predictive.

However, they are typical of the behavior of all candidate models when used in the same way. This showcases that the ability of the logistic-type model to capture asymptotic behavior, which is arguably one of its attractive properties, may become a nuisance when the upper cut-off limit is treated as a QoI during inference. The model is shown to be prone to discover asymptotes in the data and thus essentially overfit by assuming an unnecessarily restrictive parametric form. On the contrary, the arguably less expressive exponential-type model is able to capture the general trend in the observations effectively and remain robust during the forecasting stage.

\begin{figure}[htb!]
	\centering
	\includegraphics[scale=2.10]{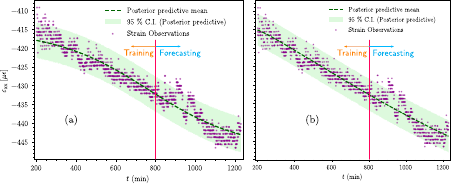}
	\caption{Posterior predictive processes using (a) the logistic-type and (b) exponential-type deterioration model. Results shown correspond to $\mathcal{M}^{(3)}$ using the entire set of observations for inference.}
	\label{Fig 13}
\end{figure}

These observations are further validated by estimating the expected utility of the candidate models. For this task, the oracle has been afforded the opportunity to train on the entire dataset, which ensures that it explains the data effectively regardless of the deterioration model form; this is illustrated by Figure \ref{Fig 13} where the posterior predictive process mean and 95\% credible intervals are plotted alongside the strain observations.
To estimate the expected utility, only the forecasting set was used for the data-based component while for the decision support-based one the probability of failure was assessed using only the last available time instance. A time-based implementation was avoided for this problem as the input-output system is linear and the model utility in terms of tracking the evolution of deterioration is effectively quantified by the data-based component.

In terms of implementation, for the data-based utility component the parameter posteriors were used to obtain strain realizations using first the corresponding deterioration model and then the observation model according to each candidate. The log-likelihood was estimated based on these predictions and the corresponding posterior realization from the prediction error standard deviation. This procedure is, in a sense, a step back from using the posterior predictive; however, since the latter operates directly on the space of observations it is not consistent with the definition provided by Eq. (\ref{Eq7}) \& (\ref{Eq8}). This is a point on which the proposed framework can be improved and the authors would like to reserve this for future research. The same procedure was used to estimate the thickness loss at the final time instant for the probability of failure calculation in the decision support-based component; also, the same Gaussian distribution was used for the yield stress.

\begin{figure}[htb!]
	\centering
	\includegraphics[scale=1.0]{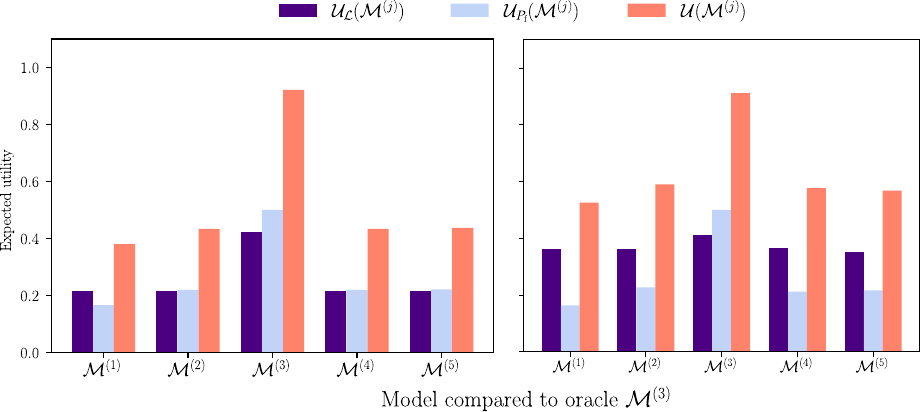}
	\caption{Expected utility of candidate models for CITL deterioration monitoring under a logistic-type (left panel) and an exponential-type (right panel) deterioration model assumption considering $\mathcal{M}^{(3)}$ as the oracle.}
	\label{Fig 14}
	
\end{figure}

Quantitatively, the expected utility is given in the form of bar charts in Figure \ref{Fig 14} for the logistic-type model (left) and  for the exponential-type model (right); in both cases equal weights have been assumed for the unified utility components. In the former, the poor selection of the deterioration model is highlighted by the fact that the individual component utilities are nearly equal, signaling a regression in the ability of the models to effectively explain the data. It is interesting to note that the increased uncertainty in predictions during the forecasting stage has caused the data-based attribute of the oracle to drop below that observed for the diagnostic task. This is a consequence of employing the log-likelihood attribute for the data-based utility definition, whereby the increased uncertainty observed in the posterior process has been effectively quantified.

In the case of the exponential-type model, expected utility estimates were unsurprisingly improved; however, in overall terms they remained at lower levels when compared to the diagnostic task. This decrease is attributed largely to the data-based component, which is affected by the increased uncertainty of the posterior process, compared to the very narrow thickness loss posteriors. Indirectly, this provides a characterization on the complexity of the problem itself, which for this task is higher-dimensional and arguably more complicated.

From a qualitative standpoint, the relative utility between the candidate models remains unaffected between tasks, which causes an earlier observation to resurface; under strict cost considerations, the decision to employ the analytical model emerges as the optimal. Returning to Occam's razor, the authors would caution against viewing this as another example reinforcing this principle. Rather, they would suggest seeing it as one that highlights the importance of Empirical Bayes.

\section{Concluding Remarks} \label{Seq 5}

In this work, the authors sought to propose comprehensive model assessment framework that can be used for different SHM tasks related to slowly evolving deterioration processes. At its core was a novel utility function definition based on weighted task-specific components that aim to assess the model in terms of its ability to fit available data but also perform downstream tasks. Such tasks are related to the use of the model as a decision support tool within the context of predictive maintenance. The framework was implemented using data from a novel experimental programme that achieves lab-scale accelerated CITL. It was assessed on different level SHM tasks of increasing complexity, ranging from system identification, to diagnostics and ultimately prognostics.

Results showed that the available models exhibited generally decreasing utilities as the tasks at hand became more complex. On the one hand, the data-based attribute proved the most robust, being undermined only in the case of poor deterioration model selection for prognostics. On the other hand, the importance of having high quality \textit{a priori} domain knowledge on the deterioration form was highlighted by the achieved decision support-based utilities. These were shown to be generally low and exhibited little variation; the latter showing that \textit{a priori} selection of higher fidelity models is a non optimal decision when lacking accurate information on the damage morphology. Ultimately, this showcases the importance of taking advantage of and incorporating historical information from deteriorated structures to design robust monitoring strategies, either for existing or future structures.

Arguably, the case study used to demonstrate the framework proposed in this work was limited in scope and restricted to laboratory conditions. Limitations in data availability due to practical restrictions somewhat narrowed down our ability to broaden the investigation of the capabilities offered by the proposed framework, especially in terms of investigating the effects of modeling inductive bias through the weights assigned to the component utilities.
However, we have provided definitions that are as general as possible, and can be easily modified and reformulated to be used effectively in other SHM applications. We view these definitions not only as a tool of appraisal, but as an ingredient in a broader, rational SHM system design framework that incorporates them alongside cost and maintenance-related utilities into a unified objective.

\section*{Data availability}

Data and code used to produce the results in this work as best as possible can be found on \url{https://github.com/nsilionis/Model-Assessment}. 

\section*{Acknowledgements}

The authors would like to gratefully acknowledge the contribution of Vaso Mantzakopoulou, who assisted substantially in organizing and executing all aspects of the laboratory tests, as well as Dora Tsiourva, who provided valuable expertise and assistance in creating the accelerated corrosive environment.




\bibliography{MSSP_References}





\end{document}